\documentclass[aps,twoside,twocolumn,nofootinbib,10pt,showpacs,floatfix]{revtex4-1}
\usepackage{amsmath,amssymb}
\usepackage{graphicx,bm}
\usepackage{slashed}
\usepackage{epstopdf}
\usepackage{ulem} 
\usepackage[usenames]{color}
\usepackage{float}
\usepackage{hyperref}
\usepackage{subfigure}
\usepackage{subfigure}
\usepackage{rotating}
\usepackage{color}
\usepackage{multirow}
\usepackage{dcolumn}
\usepackage{overpic}
\usepackage{booktabs}
\usepackage{makecell}
\usepackage{diagbox}

\renewcommand\sout{\bgroup \color{red} \ULdepth=-.5ex \ULset}

\newsavebox{\tablebox}

\begin{document}
\title{Probing hyperon CP violation from charmed baryon decays}
\author{Jian-Peng Wang$^1$}\email{wangjp20@lzu.edu.cn}
\author{Fu-Sheng Yu$^{1,2,3}$}\email{yufsh@lzu.edu.cn, corresponding author}
\affiliation{
$^1$School of Nuclear Science and Technology, Lanzhou University, Lanzhou 730000, China\\
$^2$Frontiers Science Center for Rare Isotopes, and Lanzhou Center for Theoretical Physics, Lanzhou University, Lanzhou 730000, China\\
$^3$Center for High Energy Physics, Peking University, Beijing 100871, China
}

\begin{abstract}
CP violation (CPV) in the baryon sectors has not yet been established experimentally. In this work, we propose to search for the hyperon CPV through the Cabibbo-favored charm-baryon decaying processes involving a hyperon in the final states, such as $\Lambda_c^+\to \Lambda^0\pi^+, \Lambda^0\to p\pi^-$. The CPV of the hyperon decays are purely revealed in these chain processes, since the CPV in the Cabibbo-favored charm decays are vanishing in the Standard Model (SM). The hyperons are polarized in the weak charm decays, so that we can measure the Lee-Yang asymmetric parameter $\alpha$ by the angular distributions, and then can measure the $\alpha$-induced CPV. It can be found that it is accessible for Belle II and LHCb to reach the SM prediction of the hyperon CPV in the near future. 
\end{abstract}
\maketitle

\section{Introduction}
In the modern particle physics and cosmology, one of the most crucial problems is how to understand the asymmetry between the matter and anti-matter in the Universe. It is believed that the baryogenesis mechanism can solve the problem, which requires three Sakharov criteria: baryon number violation, charge and charge-parity (CP) violation, and out of equilibrium \cite{Sakharov:1967dj}. In the Standard Model (SM) of the particle physics, CP violation (CPV) is directly generated from a weak phase in the quark mixing matrix in the Kobayashi-Maskawa (KM) mechanism \cite{Kobayashi:1973fv}. However, the SM CPV is unfortunately far smaller than the requirement of explaining the matter-dominant universe \cite{Planck:2015fie}. It implies that it is needed for a new CPV source of new physics beyond the SM. CPV have only been observed in the $K$, $B$ and $D$ meson systems \cite{Christenson:1964fg,BaBar:2001pki,Belle:2001zzw,LHCb:2019hro,PDG}, but never established in the baryon sectors yet. Besides, the visible matter of the Universe is dominantly made of baryons. Therefore, studying baryon CPV provides a good opportunity to test the SM and search for new physics.  

The strange mesons are the system to firstly observe the CP violation in 1964 \cite{Christenson:1964fg} and firstly establish the CPV  in the decay amplitudes in 1999 \cite{KTeV:1999kad}. It can be expected that the strange baryons are of good opportunity to observe the CPV in the baryon sector, since the light strange quarks are more easily produced in the experiments. The hyperon CPV was predicted at the order of $ 10^{-4}\thicksim10^{-5}$ in the SM \cite{Donoghue:1986hh,Deshpande:1994vp,Tandean:2002vy} and can be significantly enhanced to reach the order of $10^{-3}$ in some new physics models \cite{Donoghue:1986hh,Donoghue:1985ww,Chang:1994wk,He:1999bv,Chen:2001cv,Tandean:2003fr}.

The baryon CPV can be measured through the polarization of the hyperons due to their non-zero spin, in complementary to the ordinary direct CPV which measures the difference between the widths of one process and its CP conjugation. In 1957, T.D.Lee and C.N.Yang proposed the decay asymmetry parameter $\alpha$ to describe the parity violation in the $\Lambda^0\to p\pi^-$ decay, which can be measured by the angular distribution of proton in the decay of polarized hyperon \cite{Lee:1957qs}, ${\rm d}N/{\rm d}\cos\theta\propto 1+\alpha_\Lambda\cos\theta$ with $\theta$ as the angle between the direction of the hyperon polarization and the proton momentum, in analogy to the famous experiment of $\beta$ decay in the oriented Co$^{60}$ \cite{Lee:1956qn,Wu:1957my}. It can be used to measure the hyperon CPV via the so-called the $\alpha$-induced CPV, $A_{CP}^\alpha(\Lambda^0\to p\pi^-)=(\alpha_\Lambda+\bar\alpha_\Lambda)/(\alpha_\Lambda-\bar\alpha_\Lambda)$, which is different from the conventional direct CPV $A_{CP}^{\rm dir}=(\Gamma-\bar\Gamma)/(\Gamma+\bar\Gamma)$. 

High experimental precisions are required to measure the hyperon CPV due to their small values of predictions. The current most precise measurement on the CPV in the strange baryon sector is given  in the chain decay of $\Xi^-\to\Lambda^0\pi^-, \Lambda^0\to p\pi^-$ by the HyperCP collaboration \cite{HyperCP:2004zvh}, with $A_{\Xi\Lambda}^\alpha=({\alpha_\Xi\alpha_\Lambda-\bar \alpha_\Xi\bar\alpha_\Lambda})/({\alpha_\Xi\alpha_\Lambda+\bar \alpha_\Xi\bar\alpha_\Lambda})=(0.0\pm5.1\pm4.4)\times10^{-4}$. It is a combination of the CPV of $\Xi$ and $\Lambda$, $A_{\Xi\Lambda}^\alpha\approx A_{CP}^\alpha(\Xi^-\to\Lambda^0\pi^-)+A_{CP}^\alpha(\Lambda^0\to p\pi^-)$. The most precise measurement on the CPV of individual hyperon decay modes is given by the entangled hyperon anti-hyperon pair in the $J/\Psi\to \Lambda\bar\Lambda$ process by the BESIII collaboration very recently, with $A_{CP}^\alpha(\Lambda\to p\pi)=(-2.5\pm4.6\pm1.1)\times10^{-3}$ \cite{BESIII:2018cnd,BESIII:2022qax}. The CPV of $\Xi$ is recently measured in the $J/\Psi$ and $\Psi(3686)\to\Xi^-\bar\Xi^+$ processes by BESIII with much more CPV observables \cite{BESIII:2021ypr,BESIII:2022lsz}. It can be found that the $\alpha$-induced CPV is more accessible in experiments, compared to the conventional direct CPV which suffers large systematic uncertainties from the production and detection asymmetries. Note that HyperCP has been shut down for many years, and BESIII may not be able to improve the data of $J/\Psi$ by orders of magnitude. Besides, large data samples of hyperon decays will be collected at the Belle II and LHCb experiments. What is the possible method to measure the hyperon CPV with even higher precision in the near-future experiments? 

\begin{figure}[!]
\centering
\includegraphics[width=0.25\textwidth]{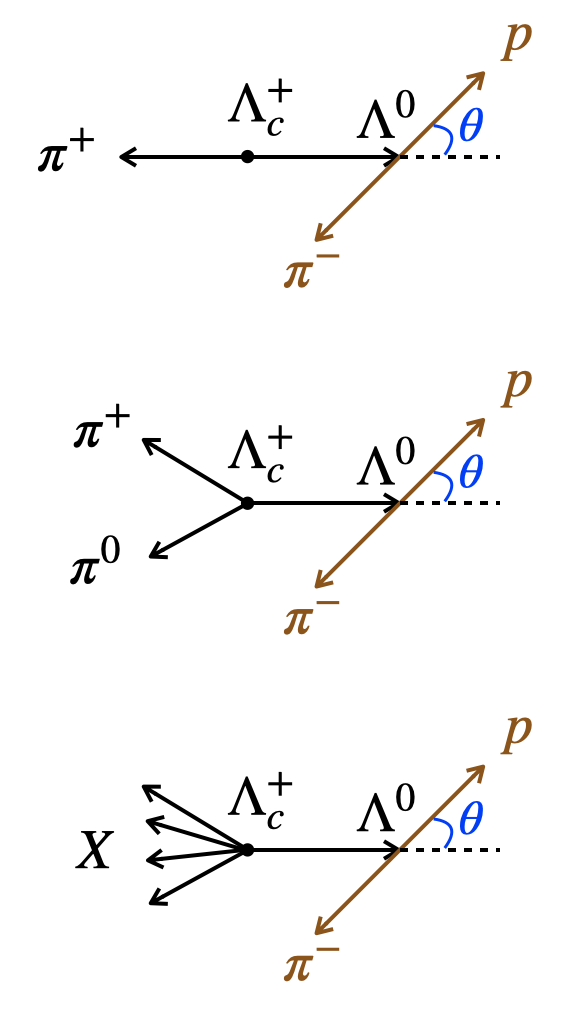}
\caption{The depicted figures of angular distributions of $\Lambda_c^+$ decays into $\Lambda^0$ and its chain decay of $\Lambda^0\to p\pi^-$. The angle $\theta$ is defined in the rest frame of $\Lambda^0$. From the top to the bottom, they are the processes of two-body decay of $\Lambda_c^+\to \Lambda^0\pi^+$, multi-body decay of $\Lambda_c^+\to \Lambda^0\pi^+\pi^0$, and the inclusive decay of $\Lambda_c^+\to \Lambda^0 X$.}\label{fig}
\end{figure}

In this work, we propose to search for the hyperon CPV through the Cabibbo-favored charm-baryon decaying processes involving a hyperon in the final states. See the depicted figures in Fig. \ref{fig}. The advantages include: (i) The CPV of the hyperon decays are purely revealed in these chain processes, since the CPV in the Cabibbo-favored charm decays are vanishing in the SM. (ii) The hyperons are polarized in the weak charm decays, so that the Lee-Yang asymmetric parameter $\alpha$ can be determined by the angular distributions, and thereby the $\alpha$-induced CPV can be measured. (iii) Compared to the direct CPV, the systematic uncertainties from the production and detection asymmetries are largely cancelled in the measurement of the $\alpha$-induced CPV. (iv) The data samples are expected to be large enough due to the large branching fractions of Cabibbo-factored charm-baryon decays. So it is accessible for Belle II and LHCb to reach the SM prediction of the hyperon CPV in the near future. 

\section{Decay asymmetry parameter $\alpha$ and the $\alpha$-induced CPV}
We briefly introduce the decay asymmetry parameter $\alpha$ which was firstly proposed by T.D.Lee and C.N.Yang in \cite{Lee:1957qs}. In the decay of completely polarized hyperon $\Lambda^0\to p\pi^-$, the angular distribution of the proton momentum in the rest frame of the hyperon, with $\theta$ the angle between the direction of the $\Lambda^0$ polarization $\vec\sigma$ and the proton momentum $\hat p$ is given as
\begin{equation}\label{eq:alpha}
	\begin{aligned}
		\frac{{\rm d}\Gamma}{{\rm d}\cos\theta}\propto 1+\alpha\cos\theta.
	\end{aligned}
\end{equation}
The parameter $\alpha$ obviously describes the parity violation in this weak-interaction process, since $\cos\theta=\vec{\sigma}\cdot\hat{p}$ is parity odd. It can be seen from the above equation that the proton is right-handed in the case of $\theta=0$, and left-handed if $\theta=\pi$. 
Therefore, the physical interpretation of $\alpha$ is the longitudinal polarization of the proton in the final state. This is a general and important property of the weak decays of baryons, which will be seen in the $\Lambda_c^+$ decays later. 

CP violation can then be measured by $\alpha$ and its $CP$ conjugation, 
\begin{equation}\label{eq:alphaCP}
	\begin{aligned}
		A_{CP}^{\alpha}\equiv\frac{\alpha-CP\alpha (CP)^{-1}}{\alpha+CP\alpha (CP)^{-1}}=\frac{\alpha+\bar{\alpha}}{\alpha-\bar{\alpha}},
	\end{aligned}
\end{equation}
where $\bar{\alpha}=C\alpha C^{-1}$, and the parity-odd property of $\alpha$ is used. It implies $\bar{\alpha}=-\alpha$ if CP is conserved. The above $A_{CP}^{\alpha}$ is called the $\alpha$-induced CPV.

There are $S$-wave and $P$-wave contributions in the amplitude $\mathcal{A}(\Lambda^0\to p\pi^-)=\bar u_p(S+P\gamma_5)u_\Lambda$. The decay asymmetry parameter is then the interference between the $S$-wave and $P$-wave amplitudes, $\alpha_\Lambda =2\mathcal{R}e(S^{*}P)/(|S|^{2}+|P|^{2})$. The $\alpha$-induced CP violation in Eq.(\ref{eq:alphaCP}) is thereby separated into two parts, $A^{\alpha}_{CP}=A-\Delta$ \cite{Donoghue:1986hh}. The first part, $A$, stems from the $S$-$P$ interference of $2\mathcal{R}e(S^{*}P)$, while the second part, $\Delta$, is from the normalization of $|S|^{2}+|P|^{2}$ which is actually the conventional direct CPV. For simplicity, the $S$-wave and $P$-wave amplitudes are expressed by two contributions with different strong phases ($\delta$) and different weak phases ($\phi$), respectively, 
$S=|S_{1}|e^{i\delta_{s,1}}e^{i\phi_{1}}+|S_{2}|e^{i\delta_{s,2}}e^{i\phi_{2}}, 
P=|P_{1}|e^{i\delta_{p,1}}e^{i\phi_{1}}+|P_{2}|e^{i\delta_{p,2}}e^{i\phi_{2}}, 
\bar{S}=-|S_{1}|e^{i\delta_{s,1}}e^{-i\phi_{1}}-|S_{2}|e^{i\delta_{s,2}}e^{-i\phi_{2}},
\bar{P}=|P_{1}|e^{i\delta_{p,1}}e^{-i\phi_{1}} +|P_{2}|e^{i\delta_{p,2}}e^{-i\phi_{2}}$.
The dependences of the CPV quantities on the relative weak and strong phases are as follows,
$A\propto \sum_{i,j=1,2}\left|S_{i}\right|\left|P_{j}\right|\sin(\delta_{p,j}-\delta_{s,i})\sin(\phi_{2}-\phi_{1})$, 
and $\Delta\propto \left|S_{1}\right|\left|S_{2}\right|\sin(\delta_{s,2}-\delta_{s,1})\sin(\phi_{2}-\phi_{1})+\left|P_{1}\right|\left|P_{2}\right|\sin(\delta_{p,2}-\delta_{p,1})\sin(\phi_{2}-\phi_{1})$. 
It can be seen that $A$ and $\Delta$ depend on the strong phases differently. 
Therefore, the $\alpha$-induced CPV is a new and complementary quantity to search for the CPV of hyperon decays, compared to the conventional direct CPV. The SM prediction in Ref.\cite{Donoghue:1986hh} shows that the direct CPV $\Delta$ is very tiny, so the $\alpha$-induced CPV is dominated by $A$.

The measurement of the $\alpha$-induced CPV has an advantage that 
the systematic uncertainties of production and detection asymmetries are largely cancelled. It can be easily understood as follows. From Eq.(\ref{eq:alpha}), the decay asymmetry parameter can also be expressed as $\alpha =\frac{\Gamma(\cos\theta>0)-\Gamma(\cos\theta<0)}{\Gamma(\cos\theta>0)+\Gamma(\cos\theta<0)}$. Then its CPV is $A_{CP}^{\alpha}\propto \Gamma(\cos\theta>0) \bar{\Gamma}(\cos\theta>0)-\Gamma(\cos\theta<0) \bar{\Gamma}(\cos\theta<0)$. This is why the current most precise measurements are all the $\alpha$-induced CPV. It will also benefit for the Belle II and LHCb experiments. 

\section{Hyperon CPV in Cabibbo-favored charmed baryon decays}
Similarly to the proton in the hyperon decay, the final-state baryons are usually polarized in the weak decays of charmed baryons. This provides a good opportunity to probe the $\alpha$-induced CPV of hyperons in the charmed baryon decays. To give an illustration, consider the Cabibbo-favored mode $\Lambda^{+}_{c}\to \Lambda^0\pi^{+}$ with $\Lambda^0\to p\pi^{-}$ as an example. The longitudinal polarization of the hyperon in this process is $\alpha(\Lambda_c\to \Lambda\pi)$ for the magnitude with the direction along the momentum of $\Lambda^0$. In principle, the polarizations of hyperon are different in different processes. We take $\alpha_{\Lambda_c}=\alpha(\Lambda_c\to \Lambda\pi)$ for simplicity here, and specify them in some other processes later.  In analogy to Eq.(\ref{eq:alpha}), the angular distribution of the proton momentum in the rest frame of $\Lambda^0$ with the angle defined in the top of Fig. \ref{fig} is
\begin{equation}\label{eq: distribution}
	\begin{aligned}
		\frac{d\Gamma}{d\cos\theta}\propto 1+\alpha_{\Lambda_c}\alpha_\Lambda \cos\theta.
	\end{aligned}
\end{equation}
It can also be derived by the helicity amplitudes. The factor of $\alpha_{\Lambda_c}$ is actually a unity in Eq.(\ref{eq:alpha}) where the hyperon is completely polarized. Then it can be used to measure the CP violation in this process. In the SM, the Cabibbo-favored charm decays are CP conserved, since there are no additional weak phases in the transition of $c\to su\bar d$. It is exactly that $\bar\alpha_{\Lambda_c}=-\alpha_{\Lambda_c}$. Therefore, the total CPV in this process is actually the one of hyperon decay, seen as
\begin{align}\label{eq:LcCPV}
A_{CP}^\alpha({\rm total})&=\frac{\alpha_{\Lambda_c}\alpha_\Lambda-\bar\alpha_{\Lambda_c}\bar\alpha_\Lambda}{\alpha_{\Lambda_c}\alpha_\Lambda+\bar\alpha_{\Lambda_c}\bar\alpha_\Lambda}\nonumber\\
&=\frac{\alpha_\Lambda+\bar\alpha_\Lambda}{\alpha_\Lambda-\bar\alpha_\Lambda}=A_{CP}^\alpha(\Lambda^0\to p\pi^-).
\end{align}
This is the main idea of this proposal that we can measure the hyperon CPV in charmed baryon decays. It is similar to the decay of $\Xi^-\to\Lambda^0\pi^-, \Lambda^0\to p\pi^-$ measured by HyperCP, but is more clear for the physical CPV source without any pollution from the primary decay. 

Except for the decay of $\Lambda^{+}_{c}\to \Lambda^0\pi^{+}$, many other Cabibbo-favored decays of charmed baryons into hyperons can be used to measure the hyperon CPV as well. Some such processes with larger branching fractions are listed in Table.\ref{tab:1}, including either $\Lambda^0$ or $\Sigma^+$ in the final states. $\Sigma^+$ weakly decays into $p\pi^0$. Many of them are recently precisely measured by the BESIII collaboration such as Refs. \cite{BESIII:2015bjk,BESIII:2015ysy}. Since only the hyperon CPV is revealed in the measurements of $\alpha$-induced CPV in these processes, we can take averages of all the $\Lambda^0$ involved modes and all the $\Sigma^+$ modes, respectively, to significantly improve the precisions. 

\begin{table}[!tbhp]
	\caption{\label{tab:test}List of some Cabibbo-favored processes of $\Lambda^{+}_{c}$ decaying into $\Lambda^0$ or $\Sigma^+$ and their branching fractions in unit of percent \cite{PDG}.}\label{tab:1} 
	\begin{tabular}{cc|cc} 
		\hline
		\hline
		$ \Lambda^{+}_{c}\rightarrow \Lambda^0 $   &\ \ \ \ BR(\%) \ \ \ \ & $ \Lambda^{+}_{c}\rightarrow \Sigma^{+} $  &\ \ \ \  BR(\%) \ \ \ \ \\ 
		\hline
		$ \Lambda^{+}_{c}\rightarrow \Lambda^0\pi^{+} $ & $ 1.30\pm0.07  $&  $\Lambda^{+}_{c}\rightarrow \Sigma^{+}\pi^0 $ & $ 1.25\pm0.10 $ \\ 
		$ \Lambda^{+}_{c}\rightarrow \Lambda^0\pi^{+}\pi^{0} $ & $ 7.1\pm0.4  $&  $\Lambda^{+}_{c}\rightarrow \Sigma^{+}\pi^{+}\pi^{-} $ & $ 4.50\pm0.25 $ \\ 
		$ \Lambda^{+}_{c}\rightarrow \Lambda^0\pi^{+}\eta $ & $ 1.84\pm0.26 $&  $\Lambda^{+}_{c}\rightarrow \Sigma^{+}\eta^{\prime} $ & $ 1.5\pm0.6 $ \\ 
		$ \Lambda^{+}_{c}\rightarrow \Lambda^0\pi^{+}\pi^{+}\pi^{-} $ & $ 3.64\pm0.29  $&  $\Lambda^{+}_{c}\rightarrow \Sigma^{+}\pi^{0}\pi^{0} $ & $ 1.55\pm0.15 $ \\
		$ \Lambda^{+}_{c}\rightarrow \Lambda^0\pi^{+}\omega $ & $ 1.5\pm0.5 $&  $\Lambda^{+}_{c}\rightarrow \Sigma^{+}\omega $ & $ 1.70\pm0.21 $\\			 
		$ \Lambda^{+}_{c}\rightarrow \Lambda^0 e^{+}\nu_{e} $ & $ 3.6\pm0.4  $&  $\rule[0pt]{0.5cm}{0.01cm} $ & $ \rule[0pt]{0.5cm}{0.01cm} $ \\
		$ \Lambda^{+}_{c}\rightarrow \Lambda^0 \mu^{+}\nu_{\mu} $ & $ 3.5\pm0.5  $&  $\rule[0pt]{0.5cm}{0.01cm}$ & $ \rule[0pt]{0.5cm}{0.01cm} $ \\
		\hline
		\hline 
	\end{tabular} 
\end{table}

It can be found that a lot of multi-body decays are useful in Table.\ref{tab:1}, with larger branching fractions. The angular distributions of the proton momentum in the rest frame of $\Lambda^0$ or $\Sigma^+$ are similar to Eq.(\ref{eq: distribution}), but with $\alpha(\Lambda_c\to\Lambda\pi)$ replaced by a quantity representing the longitudinal polarizations of the hyperons in the relevant decays of charmed baryons. See the middle figure of Fig.\ref{fig} taking $\Lambda_c^+\to \Lambda^0\pi^+\pi^0$ as an example. The dynamical calculations are usually difficult for multi-body decays, with several intermediate resonant or non-resonant contributions. No matter how large it is, there must be such a polarization in the weak decays violating the parity. In general, this polarization depends on the invariant mass of $q^2$ where $q$ is the difference between the momenta of the initial charmed baryon and the final-state hyperon, such as $q=p_{\Lambda_c}-p_\Lambda$. If we only measure the distribution of $\cos\theta$ which is defined by $\Lambda^0\to p\pi$
but un-related to $\Lambda^+_c$ decays, the polarization can be averaged. These are similar to the case of $\Lambda_c^+\to \Lambda^0\ell\nu_\ell$. By integrating out some redundant variables, the distribution function for any individual multi-body process is formulated as follows
\begin{equation}\label{eq:distribution2}
	\begin{aligned}
		\frac{d\Gamma}{d\cos\theta}\propto 1+\alpha_{\Lambda_c}^i\alpha_\Lambda \cos\theta,
	\end{aligned}
\end{equation}
where $\alpha_{\Lambda_{c}}^i =\langle\alpha_{\Lambda_{c}}^i(q^2)\rangle $ denotes the longitudinal polarization averaged over $ q^{2} $. It can also be measured for several bins of $q^2$. 
All possible resonant and non-resonant effects are included in the averaged $\alpha_{\Lambda_{c}}^i$.
Then the CPV in the multi-body Cabibbo-favored $\Lambda_c^+$ decays is the same as the two-body $\Lambda_c^+$ decay in Eq.(\ref{eq:LcCPV}), $A_{CP}^{\alpha}$(total)$=A_{CP}^{\alpha}(\Lambda\to p\pi)$. The formula in Eq.(\ref{eq:distribution2}) is more general, with $\alpha_{\Lambda_{c}}^i=\alpha_{\Lambda_c}$ in the case of two-body decays. They are the same to the $\Sigma^+$ involved modes.

\section{Discussions}
The formula of Eq.(\ref{eq: distribution}) can be easily understood for the case of un-polarized charmed baryons, while it
is also valid for polarized initial states. In fact, the baryons directly produced in the $e^+e^-$ or $pp$ collisions always have non-vanishing transverse polarizations along the normal direction of the production plane \cite{Belle:2018ttu,BESIII:2019odb,LHCb:2020iux}, no matter how small they are. Besides, the charmed baryons can be produced in the weak decays of bottom hadrons where the charmed baryons have relatively large polarizations \cite{Collaboration:2022pva}. Considering the transverse polarization of the initial charmed baryons, $P_N$, the angular distribution is 
\begin{align}\label{eq:transpol}
&{{\rm d}\Gamma\over {\rm d}\cos\theta_P~{\rm d}\cos\theta~{\rm d}\varphi}\propto 1+\alpha_{\Lambda_c}\alpha_\Lambda\cos\theta  \nonumber\\
&~~~~~~~~~+ P_N \cos\theta_P (\alpha_{\Lambda_c}+ \alpha_\Lambda\cos\theta) \\
&~~~~~~~~~ + P_N\alpha_\Lambda\sin\theta_P\sin\theta(\beta_{\Lambda_c}\sin\varphi+\gamma_{\Lambda_c}\cos\varphi),
\nonumber
\end{align}
where $\theta$, $\alpha_{\Lambda_c}$ and $\alpha_\Lambda$ are the same as defined in Eq.(\ref{eq: distribution}), $\theta_P$ is the angle between the directions of the polarization of charmed baryon and the momentum of hyperon in the rest frame of charmed baryon, $\varphi$ is the angle between the plane consisting of $P_N$ and the $\Lambda$ momentum and the decay plane of $\Lambda\to p\pi$. The parameters $\beta_{\Lambda_c}$ and $\gamma_{\Lambda_c}$ are the Lee-Yang parameters which represent the two transverse polarizations of the hyperon in the charmed baryon decays, which are not independent with each other and are usually expressed by one phase as $\beta=\sqrt{1-\alpha^2}\sin\phi$, $\gamma=\sqrt{1-\alpha^2}\cos\phi$. In the case of polarized charmed baryons, we can measure both the distributions in Eqs. (\ref{eq: distribution})  and (\ref{eq:transpol}). If integrating out the angles of $\theta_P$ and $\varphi$, the distribution in Eq. (\ref{eq:transpol}) can be simplified into the formula of Eq.  (\ref{eq: distribution}). 

If directly using distribution with the polarization of charmed baryons, $\alpha_\Lambda$ and $\bar \alpha_\Lambda$ can be seperately determined in the charmed baryon decays and their $CP$ conjugating processes respectively. Based on the $CP$ conservation law of the Cabibbo-favored charmed baryon decays, the relations between the parameters of charmed baryon decays and their charge conjugations are $\bar\alpha_{\Lambda_c}=-\alpha_{\Lambda_c}$, $\bar \beta_{\Lambda_c}=-\beta_{\Lambda_c}$ and $\bar\gamma_{\Lambda_c}=\gamma_{\Lambda_c}$, where $\alpha_{\Lambda_c}$ and $\beta_{\Lambda_c}$ are parity odd, and $\gamma_{\Lambda_c}$ are parity even. The transverse polarization of $P_N$ is somewhat complicated in the charge conjugations. In the high energy collisions, the charmed baryons are produced in two ways. One is the directly prompt production in the hadron collisions or $e^+e^-$ collisions, where the transverse polarization along the normal direction of the production plane is parity conserved. The other way of production of charmed baryons is in the weak decays of bottom hadrons where the parity is violated. If it is not possible to clearly separate the two sources of charm-baryon productions in experiment, there is no definite relation between $P_N$ and $\bar P_N$, which have to be taken as two different free parameters in the experimental analysis. If the charmed baryons are only promptly produced, the relation of $\bar P_N=P_N$ can be used. No matter on what cases, the $CP$ violation of hyperon decays can be obtained by using Eq.(\ref{eq:alphaCP}).

Although the CP violation do not depend on $\alpha_{\Lambda_c}$ in Eq.(\ref{eq: distribution}) or $\alpha_{\Lambda_{c}}^i$ in Eq.(\ref{eq:distribution2}), their values will affect the precision of measurements. The larger values  $\alpha_{\Lambda_{c}}^i$ are, the smaller errors the results of CPV are. It will be difficult to extract $ A_{CP}^{\alpha}$(total) with precision at order $ \mathcal{O}(10^{-4}\thicksim10^{-5}) $ for discovering CP violation in the hyperon, if $\alpha_{\Lambda_{c}}^i$ is tiny such that the distribution function $d\Gamma/d\cos\theta $ tends to be steady with respect to $ \cos\theta $. Although the branching fractions are given with high precisions, only a few processes of $\Lambda_c^+$ decays have been measured for the decay asymmetry parameters, such as $\alpha(\Lambda^{+}_{c}\rightarrow \Lambda^0\pi^+)=-0.84\pm0.09$, $\alpha(\Lambda^{+}_{c}\rightarrow \Lambda^0 \ell\nu_{\ell})=-0.86\pm0.04$, $\alpha(\Lambda^{+}_{c}\rightarrow \Sigma^{+}\pi^0)=-0.55\pm0.11$ \cite{PDG,Hinson:2004pj,BESIII:2019odb}. They are large enough for the measurements on the hyperon CPV. 
The large value of the $\Lambda^0 \ell\nu_{\ell}$ mode can be easily understood by the purely left-handed weak interaction for the highly boosted $\Lambda^0$ at the low $q^2$ region \cite{Li:2021qod,Geng:2020fng}.  The decay asymmetry parameter of the $\Lambda^0\pi^+$ mode is large as well, since it is dominated by the external $W$-emission diagram which is mostly factorizable with the left-handed transition of $\Lambda_c\to\Lambda$ \cite{Zou:2019kzq}. Similarly, it can be expected that the processes of $\Lambda_c^+\to \Lambda^0\pi^+\pi^0$ and $\Lambda^0\pi^+\pi^+\pi^-$ whose branching fractions are very large, might have large values of $\alpha_{\Lambda_{c}}^i$ as well due to the large factorizable contributions. Some other three-body decays of $ \Lambda^{+}_{c}$ are considered theoretically by data fitting \cite{Cen:2019ims}. Therefore, the combination of all the relevant processes might provide much more precise results on the hyperon CPV. 
On the hyperon side, the branching fractions of $ \Lambda^0\rightarrow p\pi^{-} $ and $ \Sigma^{+}\rightarrow p\pi^{0} $ are $(63.9\pm0.5)\%$ and $(51.57\pm0.30)\%$ \cite{PDG}, respectively. The decay asymmetry parameters of these two processes are $\alpha_\Lambda=-0.7542\pm0.0022$ \cite{BESIII:2022qax} and $\alpha_\Sigma=-0.982\pm0.014$ \cite{BESIII:2020fqg}, both of which are promising for the observation of CPV.

All of the above discussions can be extended into the inclusive mode $\Lambda^{+}_{c}\to \Lambda X$ whose branching fraction is as large as $(38.2^{+2.8}_{-2.2}\pm0.9)\%$\cite{BESIII:2018ciw}. From the bottom figure of Fig.\ref{fig}, the distribution function of the proton momentum in the inclusive mode is identical to the exclusive ones, with the replacement of $\alpha_{\Lambda_{c}}^i$ to a generic parameter $\mathcal{P}$ after integrating out other redundant angle variables,
\begin{equation}\label{eq:distribution3}
	\begin{aligned}
		\frac{d\Gamma}{d\cos\theta}\propto 1+\mathcal{P}\alpha_\Lambda \cos\theta.
	\end{aligned}
\end{equation}
Similarly to Eq.(\ref{eq:distribution2}), $\mathcal{P}=\langle\mathcal{P}(q^2)\rangle$ is an average of the longitudinal polarization of hyperon containing all possible effects.
In the inclusive modes, the intermediate weak decays are also included, such as $\Xi^-$ is produced in $\Lambda_c^+$ decays and then decays into $\Lambda$. 

The CP violation in the inclusive decays is more complicated than the Cabibbo-favored exclusive decay modes, since the inclusive processes include the singly Cabibbo-suppressed charmed decays where there is CPV of charmed baryons. 
Therefore, $\bar{\mathcal{P}}$ is generally un-equal to $-\mathcal{P}$. The CPV in the inclusive charmed baryon decays are 
\begin{equation}\label{eq:inclusiveCPV}
	\begin{aligned}
		A_{CP}^\alpha(\Lambda_c\to \Lambda X)=\frac{\mathcal{P}\alpha-\bar{\mathcal{P}}\bar{\alpha}}{\mathcal{P}\alpha+\bar{\mathcal{P}}\bar{\alpha}}.
	\end{aligned}
\end{equation}
Similar to the case of $A_{\Xi\Lambda}^\alpha$, it is a combination of the CP violation of both $\Lambda^{+}_{c}$ and $\Lambda$, so that we can not extract hyperon CPV independently. 
The branching fractions of the inclusive processes can be approximately taken as the sum of the branching fractions of Cabibbo-favored (CF) and singly Cabibbo-suppressed (SCS) modes, $Br(\Lambda_{c}^{+}\to \Lambda X)\approx Br({\rm CF})+Br({\rm SCS})$. CPV of charmed baryon decays only occurs in the SCS processes, which are expected  at the order of $A_{CP}^\alpha(\Lambda_{c})=\mathcal{O}(10^{-3}\thicksim10^{-4})$, similarly to the CPV of $D$ meson decays \cite{Li:2012cfa,Qin:2013tje,Li:2019hho,Saur:2020rgd}.
The CPV of inclusive charmed baryon decays are then naively estimated as $A_{CP}^\alpha(\Lambda_{c}\to \Lambda X)\approx \frac{Br({\rm SCS})}{Br(\Lambda_{c}^{+}\to \Lambda X)}\times A_{CP}^\alpha(\Lambda_{c})
		\approx \lambda^{2}A_{CP}^\alpha(\Lambda_{c})$,
where $ \lambda=0.225 $ is the Wolfenstein parameter. Consequently, $ A_{CP}^\alpha(\Lambda_{c}\to \Lambda X)$ are at the same order as the one of hyperon decays. Anyway, a non-vanishing value of Eq.(\ref{eq:inclusiveCPV}) is definitely an observation of baryon CPV.

 Experimentally, the signals of $\Lambda^{+}_{c}\to pK^{-}\pi^{+}$ has reached $1.5\times 10^{6}$ at Belle \cite{Belle:2021mvw}, which will be improved by two orders at Belle II. The considered Cabibbo-favored processes in this work have the branching fractions at the same order as that of $\Lambda^{+}_{c}\to pK^{-}\pi^{+}$. Therefore, the signal events of the $\Lambda^{+}_{c}\to \Lambda^0/\Sigma^{+}$ modes could be at the order of $10^{8}$, or even larger if combining all possible relevant processes. It can then be naively expected that the precision of the $\alpha$-induced CPV of hyperon decays could be at the order of $10^{-4}$ at Belle II, which reaches the SM predictions. There are also large data samples at LHCb since the production cross sections of charmed baryons are much larger at the hadron collisions \cite{LHCbprivate}, even though the detection effeciencies of hyperons are suppressed due to their longer lifetimes. Besides, the super tau-charm facility also has a possibility to measure the proposed hyperon CPV in charmed baryon decays \cite{Salone:2022lpt}. It can be expected that the near-future experiments of Belle II and LHCb could provide the most precise measurements on the CPV of individual process of hyperon decays, considering that the current BESIII experiment could not collect more data. 
 
 The above proposal might also be applied to the processes of $\Lambda_b^0$ decaying into $\Lambda_c^+$ via the $b\to c\bar u d$ transition, where there is no CPV in the $\Lambda_b^0$ decays and then it can be used to measure the CPV of charmed baryons. Obviously, the angular distribution, parameterization, and observable are all completely unchanged.  The experimental investigation is also practicable and promising in the future LHCb. 

\section{Summary}
In this work, we propose a novel method to measure the CP violation of hyperon decays in the Cabibbo-favored decays of charmed baryons. The main idea is that the hyperon is polarized in the weak decays of charmed baryons so we can measure the $\alpha$-induced CPV, and the pure hyperon CPV is revealed in these processes since CP is conserved in the Cabibbo-favored charm decays. Many discussions are given in detail. It is accessible that Belle II and LHCb could give the most precise measurements on the hyperon CPV in the near future. 

\vspace{0.3cm}
We are grateful to Jin-Lin Fu, Long-Ke Li, Pei-Rong Li, Xiao-Rui Lyu, Cheng-Ping Shen, Dong Xiao and Yue-Hong Xie for the useful discussions.
This work is supported by  the National Natural Science Foundation of China under Grant No. 11975112, and National Key Research and Development Program of China under Contract No. 2020YFA0406400.

\end{document}